\documentclass[sn-mathphys-ay]{sn-jnl}% Math and Physical Sciences Author Year Reference Style

\usepackage{graphicx}
\usepackage{multirow}
\usepackage{amsmath,amssymb,amsfonts}
\usepackage{amsthm}
\usepackage{mathrsfs}
\usepackage[title]{appendix}
\usepackage{xcolor}
\usepackage{textcomp}
\usepackage{manyfoot}
\usepackage{booktabs}
\usepackage{algorithm}
\usepackage{algorithmicx}
\usepackage{algpseudocode}
\usepackage{listings}
\usepackage{makecell}
\usepackage{caption}
\usepackage{placeins}
\usepackage{adjustbox}
\usepackage{booktabs}

\theoremstyle{thmstyleone}%

\theoremstyle{thmstyletwo}

\theoremstyle{thmstylethree}

\begin{document}

\title[Article Title]{Application of generalized linear models in big data: a divide and recombine (D\&R) approach}

\author*[1]{\fnm{Md. Mahadi Hassan} \sur{Nayem}}\email{mhnayem.cu.stat@outlook.com}

\author[1]{\fnm{Dr. Soma Chowdhury} \sur{Biswas}}\email{soma.stat@cu.ac.bd}

\affil*[1]{\orgdiv{Department of Statistics}, \orgname{University of Chittagong}, \orgaddress{\street{Hathazari}, \city{Chattogram}, \postcode{4331}, \country{Bangladesh}}}

\abstract{D\&R is a statistical approach designed to handle large and complex datasets. It partitions the dataset into several manageable subsets and subsequently applies the analytic method to each subset independently to obtain results. Finally, the results from each subset are combined to yield the results for the entire dataset. D\&R strategies can be implemented to fit GLMs to datasets too large for conventional methods. Several D\&R strategies are available for different GLMs, some of which are theoretically justified but lack practical validation. A significant limitation is the theoretical and practical justification for estimating combined standard errors and confidence intervals. This paper reviews D\&R strategies for GLMs and proposes a method to determine the combined standard error for D\&R-based estimators. In addition to the traditional dataset division procedures, we propose a different division method named sequential partitioning for D\&R-based estimators on GLMs. We show that the obtained D\&R estimator with the proposed standard error attains equivalent efficiency as the full data estimate. We illustrate this on a large synthetic dataset and verify that the results from D\&R are accurate and identical to those from other available R packages.}


\maketitle

\section{Introduction}\label{sec1}
Generalized linear models (GLMs) form a general framework for modeling the relationship between a response and a set of predictors. It was first introduced by \cite{nelder1972generalized} to provide a unifying set of models widely used for regression analysis. GLM is a flexible extension of ordinary OLS that enables the linear model to be related to response through a link function. Standard statistical modeling, such as GLMs, is based on a single dataset that may be very large and becomes unwieldy to work with from a computational perspective. The divide and recombine (D\&R) approach was developed to overcome these limitations in response to recent trends in data science and big data research.

The Divide and Recombine technique has proven to be an effective and generic approach to the statistical analysis of large-scale datasets\citep{guha2012large}. It partitions the data into manageable subsets, applies analytical methods to each subset, and subsequently aggregates the result. The computations are embarrassingly parallel, meaning they can run in parallel without any communication, thus facilitating efficient big-data analysis. Division in datasets can be achieved through either replicate division or conditioning variable division \citep{buhlmann2016handbook}. As per \cite{cleveland2014divide} and \cite{hafen2016divide}, replicate division uses random sampling without replacement, whereas conditioning-variable division stratifies data based on one or more variables. The replicate division enables simple and faster computations, but there may be a small loss in accuracy compared to the complete data estimate. The divisions from the dataset should be sufficiently small to be memory-manageable and amenable to further breakdown if needed.

Various compression and aggregation schemes for data cubes in linear and logistic regression models have been proposed, and their efficacy has been demonstrated. Different D\&R schemes for Poisson and multinomial models have been suggested, but their practical justification still needs to be verified. Notably, procedures for variance or standard error estimation for D\&R-based estimators for GLMs currently need to be made available. This study aims to review D\&R schemes for GLMs and proposes an aggregated standard error and confidence interval estimation scheme. The application of the D\&R strategy is demonstrated on a synthetic dataset to verify its efficiency with other standard results.

\section{D\&R for GLMs}\label{sec2}
This section provides a basic overview of different D\&R approaches for GLMs. We focus on four specific generalized linear models (GLMs): linear regression, which is used for continuous response variables; logistic regression, for binary response variables; Poisson regression, for count data; and multinomial logistic regression, which is used for categorical response variables with more than two levels.

\subsection{Multiple linear regression model}\label{subsec2.1}

For extremely large datasets to fit into memory, two practical D\&R approaches have been proposed: summary statistics D\&R and horizontal D\&R. Summary D\&R extracts relevant summary statistics that best summarize data in each subset, ensuring the aggregated estimate is as close to full data estimates as possible. \cite{lee2017sufficiency} extracted only summary data, rendering unit record data unnecessary. They proposed the procedure for the following model:

\begin{equation*}
	Y=X\beta+\epsilon,\ \epsilon \sim N(0,\sum)
\end{equation*}
With the following identity link function:
\begin{equation*}
	\theta=g\left(\mu\right)=\mu=E\left(Y\right)=X\beta
\end{equation*}
The least square estimator or ML estimator can be obtained as:
\begin{equation*}
	\hat{\beta}=\left(X^\prime X\right)^{-1} X^\prime\ Y
\end{equation*}
The summary statistics D\&R are illustrated in the following steps:
\begin{enumerate}[1.]
	\item The entire dataset is divided into s subsets, each with a similar structure, where \(X_s\) and \(Y_s\) are the vectors of the design matrix and the responses to the subset \(s\).
	
	\item Compute \(X_s^\prime X_s\) and \(X_s^\prime Y_s,\ S=1,2,3,\ldots,\ s.\)
	
	\item Recombine separate estimates using \(\left(\sum_{s=1}^{S} X_s' X_s \right)^{-1} \left(\sum_{s=1}^{S} X_s' Y_s \right)\)
	
\end{enumerate}

\subsection{Logistic regression model}\label{subsec2.2}
Let us consider a binary response variable \(y\) and the vector of covariates \(X\). Then, the logit function is defined as:
\[\ln{\left(\frac{p_i}{1-p_i}\right)}=x_i\beta\]
The estimation equation can be obtained as:
\[\frac{\delta l}{\delta\beta_j}=\ \sum_{i=1}^{n}{\left[y_i-p_i\right]x_{ij}}=0,\ j=1,2,\ldots,s\ \ \]
\[\frac{\delta l}{\delta\beta_j}=\ \sum_{i=1}^{n}{\left[y_i-\frac{e^{x_i\beta}}{1+e^{x_i\beta}}\right]x_{ij}}=0,\ j=1,2,\ldots,s\ \ \]
where \(l\) is defined as a log-likelihood function.
Now, 
\[\frac{\delta^2l}{\delta\beta_j\delta\beta_k}=\frac{\delta}{\delta\beta_k}\left(\sum_{i=1}^{n}{\left[y_i-p_i)\right]x_{ij}}\right)\]
\[H=\frac{\delta^2l}{\delta\beta_j\delta\beta_k}=\ -\sum_{i=1}^{n}{p_i\left(1-p_i\right)x_{ij}x_{ik}}\]

Here, \(H\) is the Hessian matrix that represents the second-order partial derivatives of the log-likelihood function with respect to parameters \(\beta_j\) and \(\beta_k\). 

\cite{xi2008compression} presented a compression scheme to enable high-quality aggregation of logistic regression in multidimensional data space. They retained only the model parameters and a few auxiliary measures to compress each data segment. A logistic model was constructed via an aggregation formula. The data compression method must fulfill the following criteria:  

\begin{enumerate}
	\item The compressed data in a multidimensional data cube environment should satisfy lossless or asymptotically lossless aggregation of the regression measures.
	\item ii.	As there may be many tuples in each cell, the complexity in compressed data due to space should be minimal and unaffected by the number of tuples in each cell \citep{xi2008compression}.

\end{enumerate}

The following steps are used to obtain an aggregated estimator via the D\&R technique:

\begin{enumerate}
	\item The entire dataset is divided into \(s\) subsets, each with a similar structure, where \(s=1,2,3,\ldots,S\). \(Y_s\) and \(X_s\) are vectors of responses and the design matrix.
	\item 	For the \(sth\) subset, compute the Hessian matrix:
	\[H_s=\frac{\delta^2l}{\delta\beta_j\delta\beta_k}\]
	and \({\hat{\beta}}_s\) for each subset \(s\) from the following equation: 
	\[l_s^\prime\left(\beta\right)=\frac{\delta l_s}{\delta\beta}=\sum_{i=1}^{n}{\left\{\left(y_{si}-p_{si}\right)x_{si}\ \right\}=0\ }\ \]
	\item 	Aggregate the \({\hat{\beta}}_s^\prime s\ \)  that is obtained in step 2 via the following:
	\[\hat{\beta}=\left(\sum_{s=1}^{S}H_s\right)^{-1}\ \left(\sum_{s=1}^{S}H_s\ {\hat{\beta}}_s\right)\]
\end{enumerate}

\subsection{Poisson regression model}\label{subsec2.3}
Let us consider an observed count-dependent variable \(Y\), which is assumed to follow a Poisson distribution. \(\lambda\) is called the rate parameter, which is determined by a set of k predictors \(X=\left(X_1,X_2,\ldots,X_k\right)\). Then, the expression relating to these quantities is:
\[E\left(Y=y\middle|\ X\right)=\lambda=exp({X\beta})\]
where \(X\beta\) is the systematic component and where \(exp({X\beta})\) ensures that \(\lambda\) is positive.
Therefore, the Poisson regression model for \(i\) observations can be expressed as:
\[P\left(Y_i=y_i\middle|\ X_i,\beta\right)=\frac{e^{-exp{\left(X\beta\right)}}{\exp{\left(X\beta\right)}}^{y_i}\ }{y_i!}\]
That is, the outcome variable follow a Poisson distribution for a given set of predictors with rate \(exp(X\beta)\). 
For a sample of size \(n\), the likelihood function for the Poisson regression model is given by:
\[L\left(\beta;y,X\right)=\prod_{i=1}^{n}\frac{e^{-\exp(X\beta)}{\exp{(X_i\beta)}}^{y_i}\ }{y_i!}\]
The log-likelihood function can be expressed as:
\[l\left(\beta\right)=\sum_{i=1}^{n}{[{\ X}_i\beta-\exp{(X\beta)}-\log(y_i}!)\ ]  \]

The maximization of the log-likelihood function has no close approximation. Here, techniques such as iterative weighted least squares can be implemented to estimate the regression coefficient \(\hat{\beta}\). 
The D\&R approach for poison regression described by \cite{karim2019reliability} is illustrated in the following steps:
\begin{enumerate}
	\item 	The entire dataset is divided into \(s\) subsets, each with a similar structure, where \(Y_s\) and \(X_s\) are vectors of responses and the design matrix for subset \(s\).
	\item 	For the \(sth\) partitioned subset, compute \({\hat{\beta}}_s, s=1,2,3,\ldots,S\ \)  by solving the following log-likelihood function\ \(l(\beta)\) iteratively:
	\[l\left(\beta\right)=\sum_{i=1}^{n}{[{\ X}_{si}\beta-\exp{\left(X_{si}\beta\right)}-log(y_{si}}!)\ ]\]
	\item 3.	Recombine separate estimates obtained from step 2 as: 
	\[\ \hat{\beta}=\frac{\sum_{s=1}^{S}{\hat{\beta}}_s}{S}\]
	
\end{enumerate}
\subsection{Multinomial logistic regression model}\label{subsec2.4}
In the simple logistic regression model, the response variable is a dichotomous variable taking values of $1$ and $0$ denoting success and failure as two categories. In practice, a situation arises where \(Y\) can ask to address polytomous responses, that is, \(r>2\) categories. When \(r=2\), \(Y\) is a dichotomous variable, we model the log odds that an event occurs. When \(r=2\), for a simple logistic regression model, we can form a logit function as follows:
\(logit\left(p\right)=\log{\left(\frac{p}{1-p}\right)}\)

When we deal with multi-category or polytomous response variables, i.e., \(r>2\), \(\frac{r\left(r-1\right)}{2} \) logits (odds), but only \((r-1)\) are nonredundant. These \((r-1)\) nonredundant logits can be formed in different ways, leading to different polytomous (multinomial) logistic regression models.
Let the outcome variable \(Y\) possess \(r\) categories \(Y_1=y_1\ ,\ldots\ldots\ldots\ ,Y_r=y_r\), where \(\sum_{j=1}^{r}{y_j=n}\). The probability mass function of \(Y_1,\ldots\ldots\ldots,\ Y_r\) is said to follow a multinomial distribution with probabilities \(P\left(Y_1=y_1\right)=p_1,\ P\left(Y_2=y_2\right)=p_2,\ldots\ldots,\ P\left(Y_r=y_r\right)=p_r\) as:
\[P\left[Y_1=y_1,Y_2=y_2,\ldots,Y_r=y_r\middle|\sum_{j=1}^{r}{y_j=n}\right]=\frac{\prod_{j=1}^{r}{\frac{e^{-\lambda_j}\lambda_j^{y_j}}{{\ y}_j!}\ }}{\frac{e^{-\lambda}\lambda^n}{n!}}=n!\ \prod_{j=1}^{r}{\frac{\left(\frac{{(\lambda}_j}{\lambda}\right)^{y_j}\ }{{\ y}_j!}\ }\]

which is equivalent to the multinomial form with \(\ p_j=\frac{\lambda_j}{\lambda}\)
The expression stated above can be expressed in exponential form as:
\[P\left[Y_1=y_1,Y_2=y_2,\ldots,Y_r=y_r\middle|\sum_{j=1}^{r}{y_j=n}\right]=e^{(\sum_{j=1}^{r}{y_j\ ln\left(\frac{\lambda_j}{\lambda}\right)+\ln{\left(n!\right)}-ln(}\sum_{j=1}^{r}{y_j!))\ \ \ \ }}\]

For \(Y_1,Y_2,\ldots,\ Y_r\ \), the link function is given by:
\[\ln{\left(\frac{\lambda_{ij}}{\lambda_{i1}}\right)}=x_{ij}\beta_j\ ,\ i=1,2,3,\ldots,\ n.\ \]
where, \(x_{ij}=\left(1,x_{ij1},x_{ij2},\ldots.,\ x_{ijp}\right)\) and \(\beta_j=\left(\beta_{jo},\beta_{j1},\ldots,\beta_{jp}\right)^\prime\). 
Now, the log-likelihood function is given by:
\[l=\sum_{i=1}^{n}\sum_{j=1}^{r}\left[y_{ij}\ln{\left(\frac{\lambda_{ij}}{\lambda_i}\right)}+\ln{\left(n!\right)}-\ln{\left(\sum_{j=1}^{r}{y_{ij}!}\right)}\right]\ \]
\[=\sum_{i=1}^{n}\left[\sum_{j=2}^{r}\left(y_{ij}\left(x_{ij}\beta_j\right)-\left(1+\sum_{j=1}^{r}e^{x_{ij}\beta_j}\right)+\ln{\left(n!\right)}-\ln{\left(\sum_{j=1}^{r}{y_{ij}!}\right)}\right)\ \right]\]
Therefore,
\[\frac{\delta l}{\delta\beta_j}=\sum_{i=1}^{n}{\left[y_{ij}-p_{ij}\left(\beta\right)\right]x_{ijk}=0,\ \ \ j=2,3,\ldots,r.\ \ k=0,1,2,\ldots,p\ .}\]
The information matrix is: 
\[I\left(\beta\right)=\sum_{i=1}^{n}{p_i\left(\beta\right)\left(1-p_i\left(\beta\right)\right)x_ix_i^\prime}\]
Here, \(\beta\) is an \((r-1)\) vector of parameters, and each vector consists of \((p+1)\) parameters.
The second order derivative of the log-likelihood equation in multinomial logistic regression is given by:
\[\frac{\partial^2l}{\partial\beta_j\partial\beta_k}=-\sum_{i=1}^{n}{x_{ijk}x_{ijk}^\prime p_{ij}\left(\beta\right)\left(1-p_{ij}\left(\beta\right)\right)}\]
which is the Hessian matrix.
The divide and recombine approach can be employed in large datasets with polytomous response variables. An approach described by \cite{karim2019reliability}. The steps of the D\&R approach for a generalized linear model with a log link function with a polytomous response are as follows:
\begin{enumerate}
	\item The entire dataset is divided into \(s\)  subsets, each with identical structures, where \(X_s\) and \(Y_s\) are vectors of the design matrix and responses to the subset \(s\).
	\item For the sth partitioned subset, compute \({\hat{\beta}}_s,s=1,2,3,\ldots,S\), which is obtained by solving the following equations:
	\[\frac{\delta l}{\delta\beta_{sj}}=\sum_{i=1}^{n}{\left[y_{sij}-p_{sij}\left(\beta\right)\right]x_{sijk}=0,\ \ s=1,2,3,\ldots,S,\ \ j=2,3,\ldots,r,\ \ k=0,1,2,\ldots,p\ .}\]
	\item Recombine separate estimates obtained from step 2 as:
	\[\ {\hat{\beta}}=\frac{\sum_{s=1}^{S}{\hat{\beta}}_s}{S}\]
\end{enumerate}

\section{Division and recombination procedures}\label{sec3}

According to \cite{hafen2016divide}, division in a dataset in the D\&R approach can be done through conditioning-variable division or replicate division. The condition variable division uses stratification of data based on one or more variables. , while the replicate division uses random sampling without replacement. The replicate division enables simple and faster computations, but there may be a small loss in accuracy compared to the complete data estimate. The divisions from the dataset should be manageable in memory and allow for further breakdown if necessary. 

Again, the recombination procedure includes statistical, analytic, and visual recombination \citep{guha2012large}. The statistical recombination method refers to the recombination of outputs obtained from each subset. Continued analysis of the outputs from each subset is called analytic recombination. Outputs can be handled as small data, then they result in substantial data reduction. In such cases, analyzing these small datasets is an essential component of the D\&R Approach. Visual recombination explores detailed data at their most granular level, and subsets that contain fine-grained information are utilized as highlighted by \cite{guha2009visualization}.

\subsection{Proposed division method}\label{subsec3.1}
In working with generalized linear models, our main concern is to choose the proper division method. Therefore, we followed an alternative approach.
Suppose that the dataset contains \(n\) observations. We wish to divide the dataset into \(s\) where \(s=1,2,\ldots,S\) subsets so that each subset can be fitted onto memory. Each subset contains observations \(\le\ n_s\) such that \(\sum_{s=1}^{S}n_s\ =n\). The splitting approach follows a sequential order starting from the topmost observation. These procedures are also referred to as the serial partitioning method, as described by \cite{rathi2004serial}. This procedure is different from the previously proposed division procedure for the D\&R-based estimation. We prove that this division procedure and our proposed S.E. estimation scheme provide accurate estimates of the D\&R-based estimators for GLMs.

\section{Proposed aggregated standard error estimation schemes}\label{sec4}
Due to the absence of combined standard error and confidence interval estimation for D\&R-based estimators for GLMs, this paper aims to address this gap and establish a comprehensive framework. We propose a novel method for estimating the aggregated standard error and confidence intervals. The proposed scheme is characterized by its robustness and broad applicability across different GLMs. The estimation procedure encompassed several key steps that are outlined below:

\begin{enumerate}
	\item The entire dataset is divided into \(s\) subsets.
	\item For the \(sth\) partitioned subset, compute \({\hat{\beta}}_s, s=1,2,3,\ldots,S\). 
	\item Compute:  \({V(\hat{\beta}}_s),\ s=1,2,3,\ldots,S\) for each subset.
	\item 	Now, obtain \(\frac{{V(\hat{\beta}}_s)}{S}\ \) , where S denotes the number of subsets.
	\item Obtain 
	\[Varinace =\frac{1}{S}\ \sum_{s=1}^{S}\frac{{V(\hat{\beta}}_s)}{S}\]
	\[=\frac{1}{S^2}\ \sum_{s=1}^{S}{{V(\hat{\beta}}_s)}\]
	\[=\frac{1}{S^2}\ [{V(\hat{\beta}}_1)+\ {V(\hat{\beta}}_2)+\ldots+\ {V(\hat{\beta}}_s)]\]
	\item The standard error is obtained as:
	\[S.E=\sqrt{\frac{1}{S}\ \sum_{s=1}^{S}\frac{{V(\hat{\beta}}_s)}{S}\ }\]
\end{enumerate}

\begin{figure}[h]
	\centering
	\includegraphics[width=0.9\textwidth]{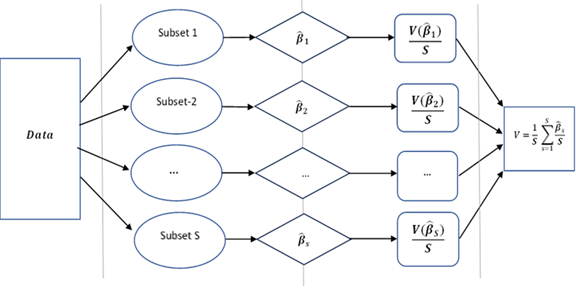}
	\caption{Proposed aggregated standard error estimation for D\&R-based estimators in GLMs}\label{Fig_1}
\end{figure}
\FloatBarrier

\section{Application to a synthetic dataset}\label{sec5}

This section evaluates the effectiveness of the D\&R approach for different GLMs. For the demonstration, we use a publicly available dataset from the UCI Machine Learning Repository, namely Heart Disease [Dataset] \citep{janosi1988uci}. Among the four datasets available at the UCI machine learning repository, we select the Cleveland Heart Disease Dataset, which has 14 variables. 

Using the synthpop package \citep{nowok2016synthpop} in R, we created a synthetic dataset via the syn() function. We used the processed Cleveland dataset, which contains only 297 rows with 14 variables. Synthpop tries to mimic the original data as much as possible, but the accuracy may vary. The proportion of participants to variables is one factor that influences the data set. Synthetic data maintains variable relationships, so more variables mean the function must account for more scenarios. With the syn() function, we set the data generation method to CART, m=1, and k=5000000. We set m as the number of synthetic datasets and k as the number of rows to generate. We generated the data one at a time and saved it. This dataset is available to download at: \href{https://www.kaggle.com/datasets/mdmahadihassannayem/cleveland-heart-disease-synthetic-dataset/data}{Cleveland Heart Disease Synthetic Dataset}

\begin{table}[h]
\caption{Attribute description of the Cleveland 
 heart disease synthetic dataset}\label{tab1}
\begin{tabular*}{\textwidth}{@{\extracolsep{\fill}}p{3.5cm}lp{2cm}l}
\toprule
Attribute name & Code & Measurement Unit & Type of Variable \\
\midrule
Age & Age & In years & Numeric \\
Sex & Sex & 0, 1 & Dichotomous \\
Chest pain type & Chest\_Pain\_Type & 1,2,3,4 & Categorical \\
Resting blood pressure (on admission to the hospital) (in mm Hg)  & Resting\_Blood\_Pressure & mm Hg & Numeric \\
Serum cholesterol (in mg/dl) & Serum\_cholesterol & mg/dl & Numeric \\
Fasting blood sugar$>$ 120 mg/dl & Fasting\_Blood\_Sugar & 0, 1 & Dichotomous \\
Resting electrocardiographic & Resting\_ECG & 0, 1, 2 & Categorical \\
Maximum heart rate achieved & Max\_Heart\_Rate\_Achieved & 71–202 & Numeric\\
Exercise-induced angina & Exercise\_Induced\_Angina & 0, 1 & Dichotomous\\
ST depression induced by exercise relative to rest & ST\_Depression\_Exercise & Depression & Numeric\\
slope: the slope of the peak exercise ST segment & Peak\_Exercise\_ST\_Segment & 0,1,2 & Categorical\\
number of major vessels (0-3) colored by flourosopy & Num\_Major\_Vessles\_Flouro & 0,1,2,3 & Count\\
A blood disorder called thalassemia & Thalassemia & 3, 6, 7 & Categorical \\
Diagnosis of Heart Disease & Diagonosis\_Heart\_Disease & 0, 1 & Dichotomous \\
\botrule
\end{tabular*}
\footnotetext{Source: \href{https://www.kaggle.com/datasets/mdmahadihassannayem/cleveland-heart-disease-synthetic-dataset/data}{Cleveland heart disease synthetic dataset}}
\end{table}
\FloatBarrier

\begin{table}[h]
	\caption{Nominal attribute description of the Cleveland heart disease synthetic dataset}\label{tab2}%
	\begin{tabular}{@{}p{6cm}l@{}}
		\toprule
		Attribute name & Description\\
		\midrule
		Sex & \makecell[{{l}}]{\raggedright 1=male\\0=female}\\
		
		Chest pain type & \makecell[{{l}}]{\raggedright 1= typical angina\\2= atypical angina\\3=nonanginal pain\\4=asymptomatic}\\
		
		Fasting blood sugar$>$120 mg/dl & \makecell[{{l}}]{\raggedright 1 = true\\0 = false}\\
		
		Resting electrocardiographic results & \makecell[{{l}}]{\raggedright 0= normal\\1= showing ST-T wave abnormality \\2= showing definite or probable left ventricular hypertrophy} \\
		
		Exercise-induced angina & \makecell[{{l}}]{\raggedright 1= yes\\0 = no}\\
		
		Slope: the slope of the peak exercise ST segment & \makecell[{{l}}]{\raggedright 1=upsloping\\2=flat\\3=downsloping}\\
		
		A blood disorder called thalassemia & \makecell[{{l}}]{\raggedright 3=normal\\6=fixed defect\\7=reversable defect}\\
		
		Diagnosis of Heart Disease & \makecell[{{l}}]{\raggedright 1 = heart disease\\ 0 = normal}\\
		\botrule
	\end{tabular}
    \footnotetext{Source: \href{https://www.kaggle.com/datasets/mdmahadihassannayem/cleveland-heart-disease-synthetic-dataset/data}{Cleveland heart disease synthetic dataset}}
\end{table}
\FloatBarrier

The synthetic dataset is systematically partitioned into ten parts using the sequential partitioning method discussed in section \ref{subsec3.1}. Each resulting subset contains 500,000 rows, with the initial subset containing the first 500,000 rows, the second subset including the subsequent 500,000 rows, and so on. The desired model is fitted independently to each subset without any communication between subsets. In fitting linear regression models, the "serum cholesterol" is designated as the response variable, whereas all other variables are treated as the predictors. For fitting the logistic regression model, the variable "diagnosis of heart disease" is treated as the response. The Poisson model utilizes the "number of major vessels colored by fluoroscopy" variable as the response, and the multinomial logistic regression model incorporates the "type of chest pain" as the response variable. The results from each subset are combined via the methods discussed in sections \ref{subsec2.1} to \ref{subsec2.4}. Standard errors and confidence intervals are calculated via the method described in section \ref{sec4}. Finally, we construct tables that show a comparison of results obtained through the D\&R approach along with the results provided by other R packages, such as stats and speedglm, to verify the accuracy of the estimates. 

\begin{sidewaystable} 
	\caption{Comparison of the coefficients of the estimates for the linear, logistic, and Poisson models using Cleveland heart disease synthetic dataset}\label{tab3} 
	\begin{tabular*}{\textheight}{@{\extracolsep\fill}lcccccccccccc} 
		\toprule
		& \multicolumn{9}{c}{Estimate} \\
		\cmidrule(lr){2-10}
		& \multicolumn{3}{c}{Linear regression model} & \multicolumn{3}{c}{Logistic regression model} & \multicolumn{3}{c}{Poisson regression model} \\
		\cmidrule(lr){2-4} \cmidrule(lr){5-7} \cmidrule(lr){8-10}
		\makecell[l]{Variables} & \makecell{Using\\ D\&R} & \makecell{Using\\ glm()} & \makecell{Using\\ speedglm()} & \makecell{Using\\ D\&R} & \makecell{Using\\ glm()} & \makecell{Using\\ speedglm()} & \makecell{Using\\ D\&R} & \makecell{Using\\ glm()} & \makecell{Using\\ speedglm()} \\
		\midrule
		(Intercept) & 160.735 & 160.735 & 160.735 & -5.027 & -5.027 & -5.027 & -3.630 & -3.630 & -3.630 \\
		
		Age & 0.824 & 0.824 & 0.824 & 0.009 & 0.009 & 0.009 & 0.047 & 0.047 & 0.047 \\
		
		Sexl & -23.482 & -23.482 & -23.482 & 0.154 & 0.154 & 0.154 & 0.214 & 0.214 & 0.214 \\
		
		Chest\_Pain\_Type2 & 7.227 & 7.227 & 7.227 & 0.218 & 0.218 & 0.218 & -0.080 & -0.080 & -0.080 \\
		
		Chest\_Pain\_Type3 & -2.251 & -2.251 & -2.251 & 0.204 & 0.204 & 0.204 & 0.147 & 0.147 & 0.147 \\
		
		Chest\_Pain\_Type 4 & 4.415 & 4.415 & 4.415 & 1.748 & 1.748 & 1.748 & 0.225 & 0.225 & 0.225 \\
		
		Resting\_Blood\_Pressure & 0.111 & 0.111 & 0.111 & 0.010 & 0.010 & 0.010 & -0.003 & -0.003 & -0.003 \\
		
		Serum\_Cholesterol & --- & --- & --- & 0.000 & 0.000 & 0.000 & 0.000 & 0.000 & 0.000 \\
		
		Fasting\_Blood\_Sugar1 & -3.026 & -3.026 & -3.026 & 0.043 & 0.043 & 0.043 & 0.460 & 0.460 & 0.460 \\
		
		Resting\_ECGl & 18.395 & 18.395 & 18.395 & -0.028 & -0.028 & -0.028 & 0.007 & 0.007 & 0.007 \\
		
		Resting\_ECG2 & 18.350 & 18.350 & 18.350 & 0.027 & 0.027 & 0.027 & -0.011 & -0.011 & -0.011 \\
		
		Max\_Heart\_Rate\_Achieved & 0.168 & 0.168 & 0.168 & 0.001 & 0.001 & 0.001 & -0.001 & -0.001 & -0.001 \\
		
		Exercise\_Induced\_Anginal & 6.375 & 6.375 & 6.375 & 0.039 & 0.039 & 0.039 & 0.028 & 0.028 & 0.028 \\
		
		ST\_Depression\_Exercise & 2.904 & 2.904 & 2.904 & 0.207 & 0.207 & 0.207 & 0.180 & 0.180 & 0.180 \\
		
		Peak\_Exercise\_ST\_Segment2 & 2.685 & 2.685 & 2.685 & 0.401 & 0.401 & 0.401 & -0.051 & -0.051 & -0.051 \\
		
		Peak\_Exercise\_ST\_Segment3 & 0.236 & 0.236 & 0.236 & -0.071 & -0.071 & -0.071 & -0.043 & -0.043 & -0.043 \\
		
		Num\_Major\_Vessels\_Flouro & -0.021 & -0.021 & -0.021 & 0.725 & 0.725 & 0.725 & --- & --- & --- \\
        
            Thalassemia6 & -11.756 & -11.756 & -11.756 & 1.947 & 1.947 & 1.947 & -0.129 & -0.129 & -0.129 \\
            
		Thalassemia7 & 2.058 & 2.058 & 2.058 & 1.751 & 1.751 & 1.751 & -0.115 & -0.115 & -0.115 \\
        
            Diagnosis\_Heart\_Disease1 & 1.353 & 1.353 & 1.353 & --- & --- & --- & 0.771 & 0.771 & 0.771 \\
		\botrule
	\end{tabular*}
\end{sidewaystable}

\begin{sidewaystable} 
	\caption{Comparison of standard errors for linear, logistic, and Poisson models using Cleveland heart disease synthetic dataset}\label{tab4} 
	\begin{tabular*}{\textheight}{@{\extracolsep\fill}lcccccccccccc} 
		\toprule
		& \multicolumn{9}{c}{Standard Error} \\
		\cmidrule(lr){2-10}
		& \multicolumn{3}{c}{Linear regression model} & \multicolumn{3}{c}{Logistic regression model} & \multicolumn{3}{c}{Poisson regression model} \\
		\cmidrule(lr){2-4} \cmidrule(lr){5-7} \cmidrule(lr){8-10}
		\makecell[l]{Variables} & \makecell{Using\\ D\&R} & \makecell{Using\\ glm()} & \makecell{Using\\ speedglm()} & \makecell{Using\\ D\&R} & \makecell{Using\\ glm()} & \makecell{Using\\ speedglm()} & \makecell{Using\\ D\&R} & \makecell{Using\\ glm()} & \makecell{Using\\ speedglm()} \\
		\midrule
		(Intercept) & 0.334 & 0.334 & 0.334 & 0.019 & 0.019 & 0.019 & 0.008 & 0.008 & 0.008 \\
		Age & 0.003 & 0.003 & 0.003 & 0.000 & 0.000 & 0.000 & 0.000 & 0.000 & 0.000 \\
		Sex1 & 0.052 & 0.052 & 0.052 & 0.003 & 0.003 & 0.003 & 0.001 & 0.001 & 0.001 \\
		Chest\_Pain\_Type2 & 0.099 & 0.099 & 0.099 & 0.005 & 0.005 & 0.005 & 0.003 & 0.003 & 0.003 \\
		Chest\_Pain\_Type3 & 0.091 & 0.091 & 0.091 & 0.005 & 0.005 & 0.005 & 0.002 & 0.002 & 0.002 \\
		Chest\_Pain\_Type4 & 0.093 & 0.093 & 0.093 & 0.005 & 0.005 & 0.005 & 0.002 & 0.002 & 0.002 \\
		Resting\_Blood\_Pressure & 0.001 & 0.001 & 0.001 & 0.000 & 0.000 & 0.000 & 0.000 & 0.000 & 0.000 \\
		Serum\_Cholesterol & --- & --- & --- & 0.000 & 0.000 & 0.000 & 0.000 & 0.000 & 0.000 \\
		Fasting\_Blood\_Sugar1 & 0.065 & 0.065 & 0.065 & 0.004 & 0.004 & 0.004 & 0.001 & 0.001 & 0.001 \\
		Resting\_ECG1 & 0.192 & 0.192 & 0.192 & 0.011 & 0.011 & 0.011 & 0.005 & 0.005 & 0.005 \\
		Resting\_ECG2 & 0.046 & 0.046 & 0.046 & 0.003 & 0.003 & 0.003 & 0.001 & 0.001 & 0.001 \\
		Max\_Heart\_Rate\_Achieved & 0.001 & 0.001 & 0.001 & 0.000 & 0.000 & 0.000 & 0.000 & 0.000 & 0.000 \\
		Exercise\_Induced\_Angina1 & 0.055 & 0.055 & 0.055 & 0.003 & 0.003 & 0.003 & 0.001 & 0.001 & 0.001 \\
		ST\_Depression\_Exercise & 0.023 & 0.023 & 0.023 & 0.001 & 0.001 & 0.001 & 0.000 & 0.000 & 0.000 \\
		Peak\_Exercise\_ST\_Segment2 & 0.054 & 0.054 & 0.054 & 0.003 & 0.003 & 0.003 & 0.001 & 0.001 & 0.001 \\
		Peak\_Exercise\_ST\_Segment3 & 0.096 & 0.096 & 0.096 & 0.005 & 0.005 & 0.005 & 0.002 & 0.002 & 0.002 \\
		Num\_Major\_Vessels\_Flouro & 0.026 & 0.026 & 0.026 & 0.001 & 0.001 & 0.001 & --- & --- & --- \\
		Thalassemia6 & 0.100 & 0.100 & 0.100 & 0.005 & 0.005 & 0.005 & 0.002 & 0.002 & 0.002 \\
		Thalassemia7 & 0.056 & 0.056 & 0.056 & 0.003 & 0.003 & 0.003 & 0.001 & 0.001 & 0.001 \\
		Diagnosis\_Heart\_Disease1 & 0.057 & 0.057 & 0.057 & --- & --- & --- & 0.001 & 0.001 & 0.001 \\
		
		\botrule
	\end{tabular*}
\end{sidewaystable}

\begin{sidewaystable}
	\caption{Comparison of t values and z values for linear, logistic, and Poisson models using Cleveland heart disease synthetic dataset}\label{tab5}
	\centering
	\begin{tabular*}{\textwidth}{@{\extracolsep{\fill}}lccccccccc}
		\toprule
		& \multicolumn{3}{c}{t value} & \multicolumn{6}{c}{z value} \\
		\cmidrule(lr){2-4} \cmidrule(lr){5-10}
		& \multicolumn{3}{c}{Linear regression model} & \multicolumn{3}{c}{Logistic regression model} & \multicolumn{3}{c}{Poisson regression model} \\
		\cmidrule(lr){2-4} \cmidrule(lr){5-7} \cmidrule(lr){8-10}
		\makecell[l]{Variables} & \makecell{Using\\ D\&R} & \makecell{Using\\ glm()} & \makecell{Using\\ speedglm()} & \makecell{Using\\ D\&R} & \makecell{Using\\ glm()} & \makecell{Using\\ speedglm()} & \makecell{Using\\ D\&R} & \makecell{Using\\ glm()} & \makecell{Using\\ speedglm()} \\
		
		\midrule
		(Intercept) & 481.186 & 481.193 & 481.193 & -271.467 & -271.494 & -271.494 & -427.110 & -427.123 & -427.123 \\
		Age & 281.828 & 281.832 & 281.832 & 58.634 & 58.637 & 58.637 & 642.566 & 642.583 & 642.583 \\
		Sex 1 & -452.615 & -452.620 & -452.620 & 52.900 & 52.902 & 52.902 & 158.612 & 158.619 & 158.619 \\
		Chest\_Pain\_Type2 & 72.804 & 72.805 & 72.805 & 40.626 & 40.638 & 40.638 & -27.539 & -27.535 & -27.535 \\
		Chest\_Pain\_Type3 & -24.715 & -24.715 & -24.716 & 42.833 & 42.846 & 42.846 & 61.962 & 61.956 & 61.956 \\
		Chest\_Pain\_Type4 & 47.519 & 47.520 & 47.520 & 366.039 & 366.093 & 366.093 & 96.500 & 96.499 & 96.499 \\
		Resting\_Blood\_Pressure & 80.943 & 80.944 & 80.944 & 139.919 & 139.935 & 139.935 & -82.761 & -82.763 & -82.763 \\
		Serum\_Cholesterol & --- & --- & --- & 16.333 & 16.336 & 16.336 & 35.663 & 35.663 & 35.663 \\
		Fasting\_Blood\_Sugar1 & -46.749 & -46.750 & -46.750 & 12.103 & 12.102 & 12.102 & 346.776 & 346.778 & 346.778 \\
		Resting\_ECG1 ${ }^{-1}$ & 95.654 & 95.664 & 95.664 & -2.626 & -2.631 & -2.631 & 1.455 & 1.478 & 1.478 \\
		Resting\_ECG2 & 402.510 & 402.512 & 402.512 & 10.860 & 10.860 & 10.860 & -9.941 & -9.936 & -9.936 \\
		Max\_Heart\_Rate\_Achieved & 135.464 & 135.465 & 135.466 & 15.128 & 15.128 & 15.128 & -32.403 & -32.391 & -32.391 \\
		Exercise\_Induced\_Angina 1 & 115.704 & 115.705 & 115.705 & 13.301 & 13.301 & 13.301 & 23.309 & 23.316 & 23.316 \\
		ST\_Depression\_Exercise & 125.208 & 125.210 & 125.210 & 161.523 & 161.532 & 161.532 & 377.397 & 377.412 & 377.412 \\
		Peak\_Exercise\_ST\_Segment2 & 50.066 & 50.066 & 50.066 & 141.418 & 141.430 & 141.430 & -38.248 & -38.249 & -38.249 \\
		Peak\_Exercise\_ST\_Segment3 & 2.446 & 2.446 & 2.446 & -13.655 & -13.656 & -13.656 & -20.130 & -20.126 & -20.126 \\
		Num\_Major\_Vessels\_Flouro & -0.813 & -0.813 & -0.813 & 500.713 & 500.773 & 500.773 & --- & --- & --- \\
        
		Thalassemia6 & -118.147 & -118.150 & -118.150 & 365.542 & 365.598 & 365.598 & -60.752 & -60.748 & -60.748 \\
        
		Thalassemia 7 & 36.908 & 36.908 & 36.908 & 648.015 & 648.086 & 648.086 & -87.304 & -87.317 & -87.317 \\
        
		Diagnosis\_Heart\_Disease1 & 23.665 & 23.665 & 23.665 & --- & --- & --- & 534.456 & 534.462 & 534.462 \\
		
		\botrule
	\end{tabular*}
\end{sidewaystable}

\begin{sidewaystable}
	\caption{Comparison of P values for or linear, logistic, and Poisson models using Cleveland heart disease synthetic dataset}\label{tab6}
	\centering
	\begin{tabular*}{\textwidth}{@{\extracolsep{\fill}}lccccccccc}
		\toprule
		& \multicolumn{3}{c}{Pr $(>|t| )$} & \multicolumn{6}{c}{Pr $(>|z| )$} \\
		\cmidrule(lr){2-4} \cmidrule(lr){5-10}
		& \multicolumn{3}{c}{Linear regression model} & \multicolumn{3}{c}{Logistic regression model} & \multicolumn{3}{c}{Poisson regression model} \\
		\cmidrule(lr){2-4} \cmidrule(lr){5-7} \cmidrule(lr){8-10}
		\makecell[l]{Variables} 
		& \makecell{Using\\ D\&R} & \makecell{Using\\ glm()} & \makecell{Using\\ speedglm()} & \makecell{Using\\ D\&R} & \makecell{Using\\ glm()} & \makecell{Using\\ speedglm()} & \makecell{Using\\ D\&R} & \makecell{Using\\ glm()} & \makecell{Using\\ speedglm()} \\
		
\midrule
(Intercept) & 0.000 & $<2e-16$ & 0.000 & 0.000 & $<2e-16$ & 0.000 & 0.000 & $<2e-16$ & 0.000 \\
Age & 0.000 & $<2e-16$ & 0.000 & 0.000 & $<2e-16$ & 0.000 & 0.000 & $<2e-16$ & 0.000 \\
Sex1 & 0.000 & $<2e-16$ & 0.000 & 0.000 & $<2e-16$ & 0.000 & 0.000 & $<2e-16$ & 0.000 \\
Chest\_Pain\_Type2 & 0.000 & $<2e-16$ & 0.000 & 0.000 & $<2e-16$ & 0.000 & 0.000 & $<2e-16$ & 0.000 \\
Chest\_Pain\_Type3 & 0.000 & $<2e-16$ & 0.000 & 0.000 & $<2e-16$ & 0.000 & 0.000 & $<2e-16$& 0.000 \\
Chest\_Pain\_Type4 & 0.000 & $<2e-16$ & 0.000 & 0.000 & $<2e-16$ & 0.000 & 0.000 & $<2e-16$ & 0.000 \\
Resting\_Blood\_Pressure & 0.000 & $<2e-16$ & 0.000 & 0.000 & $<2e-16$ & 0.000 & 0.000 & $<2e-16$ & 0.000 \\
Serum\_Cholesterol & --- & --- & --- & 0.000 & $<2e-16$ & 0.000 & 0.000 & $<2e-16$ & 0.000 \\ 
Fasting\_Blood\_Sugar1 & 0.000 & $<2e-16$ & 0.000 & 0.000 & $<2e-16$ & 0.000 & 0.000 & $<2e-16$ & 0.000 \\
Resting\_ECG1 & 0.000 & $<2e-16$ & 0.000 & 0.009 & 0.009 & 0.009 & 0.000 & $<2e-16$ & 0.000 \\
Resting\_ECG2 & 0.000 & $<2e-16$ & 0.000 & 0.000 & $<2e-16$ & 0.000 & 0.142 & 0.139 & 0.139 \\
Max\_Heart\_Rate\_Achieved & 0.000 & $<2e-16$ & 0.000 & 0.000 & $<2e-16$ & 0.000 & 0.000 &$<2e-16$& 0.000 \\
Exercise\_Induced\_Anginal & 0.000 & $<2e-16$ & 0.000 & 0.000 & $<2e-16$& 0.000 & 0.000 & $<2e-16$ & 0.000 \\
ST\_Depression\_Exercise & 0.000 & $<2e-16$ & 0.000 & 0.000 & $<2e-16$ & 0.000 & 0.000 & $<2e-16$ & 0.000 \\
Peak\_Exercise\_ST\_Segment2 & 0.000 & $<2e-16$ & 0.000 & 0.000 & $<2e-16$ & 0.000 & 0.000 & $<2e-16$ & 0.000 \\
Peak\_Exercise\_ST\_Segment3 & 0.014 & 0.014 & 0.014 & 0.000 & $<2e-16$ & 0.000 & 0.000 & $<2e-16$ & 0.000 \\
Num\_Major\_Vessels\_Flouro & 0.416 & 0.416 & 0.416 & 0.000 & $<2e-16$ & 0.000 & --- & --- & ---\\
Thalassemia6 & 0.000 & $<2e-16$ & 0.000 & 0.000 & $<2e-16$ & 0.000 & 0.000 & $<2e-16$ & 0.000 \\
halassemia7 & 0.000 & $<2e-16$ & 0.000 & 0.000 & $<2e-16$ & 0.000 & 0.000 & $<2e-16$ & 0.000 \\
Diagnosis\_Heart\_Disease1 & 0.000 & $<2e-16$ & 0.000 &  ---  & --- & --- & 0.000 & $<2e-16$ & 0.000 \\
		
\botrule
\end{tabular*}
\end{sidewaystable}

\begin{sidewaystable}
	\caption{Comparison of confidence intervals for linear, logistic, and Poisson models using Cleveland heart disease synthetic dataset}\label{tab7} 
	\begin{tabular*}{\textheight}{@{\extracolsep\fill}lcccccccccccc} 
		\toprule
		& \multicolumn{9}{c}{Confidence Interval} \\
		\cmidrule(lr){2-10}
		& \multicolumn{3}{c}{Linear regression model} & \multicolumn{3}{c}{Logistic regression model} & \multicolumn{3}{c}{Poisson regression model} \\
		\cmidrule(lr){2-4} \cmidrule(lr){5-7} \cmidrule(lr){8-10}
		\makecell[l]{Variables} 
		& \makecell{Using\\ D\&R} & \makecell{Using\\ glm()} & \makecell{Using\\ speedglm()} & \makecell{Using\\ D\&R} & \makecell{Using\\ glm()} & \makecell{Using\\ speedglm()} & \makecell{Using\\ D\&R} & \makecell{Using\\ glm()} & \makecell{Using\\ speedglm()} \\
\midrule
\tiny(Intercept) & \tiny [160.08, 161.39] & \tiny [160.08, 161.39] & \tiny [160.08, 161.39] & \tiny [-5.06,-4.99] & \tiny [-5.06, -4.99] & \tiny [-5.06,-4.99] & \tiny [-3.65, -3.61] & \tiny [-3.65, -3.61] & \tiny [-3.65, -3.61] \\

\tiny Age & \tiny [0.82,0.83] & \tiny [0.82,0.83] & \tiny [0.82,0.83] & \tiny [0.01,0.01] & \tiny [0.01,0.01] & \tiny [0.01,0.01] & \tiny [0.05,0.05] & \tiny [0.05,0.05] & \tiny [0.05,0.05] \\

\tiny Sex1 & \tiny [-23.58,-23.38] & \tiny [-23.58, -23.38] & \tiny [-23.58,-23.38] & \tiny [0.15, 0.16] & \tiny [0.15, 0.16] & \tiny [0.15, 0.16] & \tiny [0.21,0.22] & \tiny [0.21,0.22] & \tiny [0.21,0.22] \\

\tiny Chest\_Pain\_Type2 & \tiny [7.03, 7.42] & \tiny [7.03, 7.42] & \tiny [7.03, 7.42] & \tiny [0.21,0.23] & \tiny [0.21,0.23] & \tiny [0.21,0.23] & \tiny [-0.09,-0.07] & \tiny [-0.09,-0.07] & \tiny [-0.09,-0.07] \\

\tiny Chest\_Pain\_Type3 & \tiny [-2.43, -2.07] & \tiny [-2.43,-2.07] & \tiny [-2.43, -2.07] & \tiny [0.19, 0.21] & \tiny [0.19, 0.21] & \tiny [0.19, 0.21] & \tiny [0.14,0.15] & \tiny [0.14,0.15] & \tiny [0.14,0.15] \\

\tiny Chest\_Pain\_Type 4 & \tiny [4.23,4.6] & \tiny [4.23,4.6] & \tiny [4.23,4.6] & \tiny [1.74, 1.76] & \tiny [1.74, 1.76] & \tiny [1.74, 1.76] & \tiny [0.22,0.23] & \tiny [0.22,0.23] & \tiny [0.22,0.23] \\

\tiny Resting\_Blood\_Pressure & \tiny [0.11, 0.11] & \tiny [0.11, 0.11] & \tiny [0.11, 0.11] & \tiny [0.01, 0.01] & \tiny [0.01, 0.01] & \tiny [0.01, 0.01] & \tiny [0,0] & \tiny [0,0] & \tiny [0,0] \\

\tiny Serum\_Cholesterol & \tiny --- & \tiny --- & \tiny --- & \tiny [0,0] & \tiny [0,0] & \tiny [0,0] & \tiny [0,0] & \tiny [0,0] & \tiny [0,0] \\

\tiny Fasting\_Blood\_Sugar 1 & \tiny [-3.15,-2.9] & \tiny [-3.15, -2.9] & \tiny [-3.15,-2.9] & \tiny [0.04, 0.05] & \tiny [0.04, 0.05] & \tiny [0.04, 0.05] & \tiny [0.46,0.46] & \tiny [0.46,0.46] & \tiny [0.46,0.46] \\

\tiny Resting\_ECG1 & \tiny [18.02, 18.77] & \tiny [18.02, 18.77] & \tiny [18.02, 18.77] & \tiny [-0.05,-0.01] & \tiny [-0.05, -0.01] & \tiny [-0.05, -0.01] & \tiny [0,0.02] & \tiny [0,0.02] & \tiny [0,0.02] \\

\tiny Resting\_ECG2 & \tiny [18.26, 18.44] & \tiny [18.26, 18.44] & \tiny [18.26, 18.44] & \tiny [0.02,0.03] & \tiny [0.02, 0.03] & \tiny [0.02,0.03] & \tiny [-0.01,-0.01] & \tiny [-0.01,-0.01] & \tiny [-0.01,-0.01] \\

\tiny Max\_Heart\_Rate\_Achieved & \tiny [0.17, 0.17] & \tiny [0.17, 0.17] & \tiny [0.17, 0.17] & \tiny [0,0] & \tiny [0,0] & \tiny [0,0] & \tiny [0,0] & \tiny [0,0] & \tiny [0,0] \\

\tiny Exercise\_Induced\_Anginal & \tiny [6.27,6.48] & \tiny [6.27,6.48] & \tiny [6.27,6.48] & \tiny [0.03, 0.04] & \tiny [0.03, 0.04] & \tiny [0.03, 0.04]  & \tiny [0.03,0.03] & \tiny [0.03,0.03] & \tiny [0.03,0.03] \\

\tiny ST\_Depression\_Exercise & \tiny [2.86, 2.95] & \tiny [2.86, 2.95] & \tiny [2.86, 2.95] & \tiny [0.2,0.21] & \tiny [0.2,0.21] & \tiny [0.2, 0.21] &  \tiny [0.18,0.18] & \tiny [0.18,0.18] & \tiny [0.18,0.18] \\

\tiny Peak\_Exercise\_ST\_Segment2 & \tiny [2.58, 2.79] & \tiny [2.58, 2.79] & \tiny [2.58, 2.79] & \tiny [0.4, 0.41] & \tiny [0.4, 0.41] & \tiny [0.4, 0.41] & \tiny [-0.05,-0.05] & \tiny [-0.05,-0.05] & \tiny [-0.05, -0.05] \\

\tiny Peak\_Exercise\_ST\_Segment 3 & \tiny [0.05, 0.42] & \tiny [0.05, 0.42] & \tiny [0.05, 0.42] & \tiny [-0.08, -0.06] & \tiny [-0.08, -0.06] & \tiny [-0.08,-0.06] & \tiny [-0.05,-0.04] & \tiny [-0.05,-0.04] & \tiny [-0.05, -0.04]\\

\tiny Num\_Major\_Vessels\_Flouro & \tiny [-0.07, 0.03] & \tiny [-0.07, 0.03] &\tiny [-0.07, 0.03] & \tiny [0.72,0.73] & \tiny [0.73,0.73] & \tiny [0.72, 0.73] & \tiny --- & \tiny --- & \tiny --- \\

\tiny Thalassemia6 & \tiny [-11.95,-11.56] & \tiny [-11.95,-11.56] & \tiny [-11.95, -11.56]& \tiny [1.94, 1.96] & \tiny [1.94, 1.96] & \tiny [1.94, 1.96] & \tiny [-0.13,-0.12] & \tiny [-0.13,-0.12] & \tiny [-0.13, -0.12] \\

\tiny Thalassemia7 & \tiny [1.95, 2.17] & \tiny [1.95, 2.17] & \tiny [1.95, 2.17] & \tiny [1.75, 1.76] & \tiny [1.75, 1.76] & \tiny [1.75, 1.76] & \tiny [-0.12,-0.11] & \tiny [-0.12, -0.11] & \tiny [-0.12, -0.11] \\

\tiny Diagnosis\_Heart\_Diseasel & \tiny [1.24, 1.47] & \tiny [1.24, 1.47] & \tiny [1.24, 1.47] & \tiny --- & \tiny --- & \tiny --- & \tiny [0.77,0.77] & \tiny [0.77,0.77] & \tiny [0.77,0.77] \\

\bottomrule
\end{tabular*}
\end{sidewaystable}

\begin{sidewaystable}
	\caption{Comparison of the Coefficients of the Estimate and Standard Errors for the multinomial logistic regression models using Cleveland heart disease synthetic dataset}\label{tab8}
	\centering
	\begin{tabular*}{\textwidth}{@{\extracolsep{\fill}}lcccccccccccc}
		\toprule
		& \multicolumn{6}{c}{Estimate} & \multicolumn{4}{c}{Standard Error} \\
		\cmidrule(lr){2-7} \cmidrule(lr){8-13}
		\makecell[l]{Variables} 
		& \multicolumn{2}{c}{Estimate 2} & \multicolumn{2}{c}{Estimate 3} & \multicolumn{2}{c}{Estimate 4} 
		& \multicolumn{2}{c}{Standard Error 2} & \multicolumn{2}{c}{Standard Error 3} & \multicolumn{2}{c}{Standard Error 4}\\
		\cmidrule(lr){2-3} \cmidrule(lr){4-5} \cmidrule(lr){6-7} \cmidrule(lr){8-9} \cmidrule(lr){10-11} \cmidrule(lr){12-13}
		& \makecell{\tiny Using\\ \tiny D\&R} & \makecell{\tiny Using\\ \tiny multinom()} 
		& \makecell{\tiny Using\\ \tiny D\&R} & \makecell{Using\\ \tiny multinom()} 
		& \makecell{\tiny Using\\ \tiny D\&R} & \makecell{Using\\ \tiny multinom()} 
		& \makecell{\tiny Using\\ \tiny D\&R} & \makecell{Using\\ \tiny multinom()} 
		& \makecell{\tiny Using\\ \tiny D\&R} & \makecell{Using\\ \tiny multinom()}
		& \makecell{\tiny Using\\ \tiny D\&R} & \makecell{Using\\ \tiny multinom()} \\
		
		\midrule
		
		\tiny (Intercept) & \tiny 7.865 & \tiny 7.865 & \tiny 10.760 & \tiny 10.759 & \tiny 12.482 & \tiny 12.482 & \tiny 0.034 & \tiny 0.034 & \tiny 0.031 & \tiny 0.031 & \tiny 0.032 & \tiny 0.032 \\
		\tiny Age & \tiny -0.030 & \tiny -0.030 & \tiny -0.027 & \tiny -0.027 & \tiny -0.044 & \tiny -0.044 & \tiny 0.000 & \tiny 0.000 & \tiny 0.000 & \tiny 0.000 & \tiny 0.000 & \tiny 0.000 \\
		\tiny Sex1 & \tiny -0.798 & \tiny -0.798 & \tiny -1.254 & \tiny -1.254 & \tiny -1.265 & \tiny -1.265 & \tiny 0.005 & \tiny 0.005 & \tiny 0.005 & \tiny 0.005 & \tiny 0.005 & \tiny 0.005 \\
		\tiny Resting\_Blood\_Pressure & \tiny -0.021 & \tiny -0.021 & \tiny -0.027 & \tiny -0.027 & \tiny -0.023 & \tiny -0.023 & \tiny 0.000 & \tiny 0.000 & \tiny 0.000 & \tiny 0.000 & \tiny 0.000 & \tiny 0.000 \\
		\tiny Serum\_Cholesterol & \tiny 0.002 & \tiny 0.002 & \tiny -0.002 & \tiny -0.002 & \tiny 0.001 & \tiny 0.001 & \tiny 0.000 & \tiny 0.000 & \tiny 0.000 & \tiny 0.000 & \tiny 0.000 & \tiny 0.000 \\
		\tiny Fasting\_Blood\_Sugar1 & \tiny -0.835 & \tiny -0.835 & \tiny -0.050 & \tiny -0.050 & \tiny -1.016 & \tiny -1.016 & \tiny 0.006 & \tiny 0.006 & \tiny 0.005 & \tiny 0.005 & \tiny 0.005 & \tiny 0.005 \\
		\tiny Resting\_ECG1 & \tiny 0.117 & \tiny 0.116 & \tiny -0.062 & \tiny \tiny -0.062 & \tiny 0.012 & \tiny 0.012 & \tiny 0.007 & \tiny 0.007 & \tiny 0.006 & \tiny 0.006 & \tiny 0.007 & \tiny 0.007 \\
		\tiny Resting\_ECG2 & \tiny -0.838 & \tiny -0.838 & \tiny -0.483 & \tiny \tiny -0.483 & \tiny -0.133 & \tiny -0.133 & \tiny 0.004 & \tiny 0.004 & \tiny 0.004 & \tiny 0.004 & \tiny 0.004 & \tiny 0.004 \\
		\tiny Max\_Heart\_Rate\_Achieved & \tiny -0.008 & \tiny -0.008 & \tiny -0.016 & \tiny -0.016 & \tiny -0.034 & \tiny -0.034 & \tiny 0.000 & \tiny 0.000 & \tiny 0.000 & \tiny 0.000 & \tiny 0.000 & \tiny 0.000 \\
		\tiny Exercise\_Induced\_Anginal & \tiny 0.095 & \tiny 0.095 & \tiny 0.343 & \tiny 0.343 & \tiny 1.862 & \tiny 1.862 & \tiny 0.007 & \tiny 0.007 & \tiny 0.006 & \tiny 0.006 & \tiny 0.006 & \tiny 0.006 \\
		\tiny ST\_Depression\_Exercise & \tiny -0.872 & \tiny -0.872 & \tiny -0.172 & \tiny -0.172 & \tiny -0.365 & \tiny -0.365 & \tiny 0.003 & \tiny 0.003 & \tiny 0.002 & \tiny 0.002 & \tiny 0.002 & \tiny 0.002 \\
		\tiny Peak\_Exercise\_ST\_Segment2 & \tiny -0.493 & \tiny -0.493 & \tiny -0.422 & \tiny -0.422 & \tiny -0.633 & \tiny -0.632 & \tiny 0.005 & \tiny 0.005 & \tiny 0.004 & \tiny 0.004 & \tiny 0.005 & \tiny 0.005 \\
		\tiny Peak\_Exercise\_ST Segment3 & \tiny -0.565 & \tiny -0.565 & \tiny -0.457 & \tiny -0.457 & \tiny -0.545 & \tiny -0.545 & \tiny 0.009 & \tiny 0.009 & \tiny 0.007 & \tiny 0.007 & \tiny 0.008 & \tiny 0.008 \\
		\tiny Num\_Major\_Vessels\_Flouro & \tiny -0.042 & \tiny -0.042 & \tiny 0.123 & \tiny 0.123 & \tiny 0.212 & \tiny 0.212 & \tiny 0.003 & \tiny 0.003 & \tiny 0.002 & \tiny 0.002 & \tiny 0.002 & \tiny 0.002 \\
		\tiny Thalassemia6 & \tiny 0.203 & \tiny 0.203 & \tiny 0.209 & \tiny 0.209 & \tiny 0.344 & \tiny 0.344 & \tiny 0.011 & \tiny 0.011 & \tiny 0.009 & \tiny 0.009 & \tiny 0.009 & \tiny 0.009 \\
		\tiny Thalassemia7 & \tiny -0.034 & \tiny -0.035 & \tiny -0.044 & \tiny -0.044 & \tiny 0.385 & \tiny 0.385 & \tiny 0.005 & \tiny 0.005 & \tiny 0.005 & \tiny 0.005 & \tiny 0.005 & \tiny 0.005 \\
		\tiny Diagnosis\_Heart\_Disease1 & \tiny 0.105 & \tiny 0.105 & \tiny 0.153 & \tiny 0.153 & \tiny 1.675 & \tiny 1.675 & \tiny 0.005 & \tiny 0.005 & \tiny 0.005 & \tiny 0.005 & \tiny 0.005 & \tiny 0.005 \\
		
		\bottomrule
	\end{tabular*}
\end{sidewaystable}
\FloatBarrier

The coefficient of estimates of linear, logistic, Poisson, and multinomial logistic regression, obtained via D\&R, glm(), and speedglm(), is presented in Table \ref{tab3}. The estimates obtained using the D\&R method align precisely with those obtained from glm() and speedglm() functions in R. Table \ref{tab4} presents the standard errors for different models obtained via the method discussed in section \ref{sec4}. The standard deviations obtained through D\&R mirror those produced by glm() and speedglm(). Table \ref{tab5},\ref{tab6},\ref{tab7} presents the t and z values, probability values, and confidence intervals for different GLMs under consideration. D\&R exhibits exact and identical results with those obtained from glm() and speedglm(). Furthermore, Table \ref{tab8} illustrates the coefficients of estimates and standard errors for multinomial logistic regression models employing both D\&R and multinom() from the nnet package in R and no significant mismatch is seen among results. We can infer that the D\&R methods discussed in the preceding sections, in conjunction with the division approach and the proposed standard error estimation techniques, produce accurate estimates as glm() and speedglm() functions. A noticeable reduction in object and reduced computation time is seen upon fitting the models in the D\&R strategy. This verifies the efficiency of the D\&R method for fitting GLMs in large datasets, particularly in terms of computational time and memory efficiency.

\section{Conclusion}\label{sec6}

This paper presents an extensive overview of various divide and recombine (D\&R) techniques for different GLMs to large-scale datasets. The preliminary section focuses on different divide and recombine approaches with their operational principles for linear regression models. We detail the D\&R framework with respect to logistic, Poisson, and multinomial logistic regression models. The next part of the paper elaborates on current division strategies; then we introduce a new division strategy called sequential partitioning for D\&R-based estimators for GLMs. We also present a novel method to derive the standard errors and confidence intervals for D\&R-based estimators, which enable inference on variable estimates within various models. We then created a synthetic dataset of 5M observations and 14 variables to evaluate the performance of D\&R methods. Using this dataset, we estimated linear, logistic, Poisson, and multinomial logistic regression models using the D\&R approach. The resulting outcomes were benchmarked against the results from glm () and speedglm () in R, showing no significant differences. This provides confirmation of methodological correctness as well as correct implementation of D\&R methods in GLMs for large data sets.

\backmatter

\bmhead{Acknowledgements}
I would like to thank my thesis supervisor, Dr. Soma Chowdhury Biswas, for her guidance and support throughout this research. I am grateful to the Department of Statistics, University of Chittagong, for providing me with the resources. This work was derived from my M.S. thesis, and I extend my heartfelt appreciation to my family and peers for their encouragement throughout this journey.

\section*{Declarations}

\begin{itemize}
	\item Funding: This study received no funding.
	\item Conflict of interest: The authors declare that they have no affiliations with or involvement in any organization or entity with any financial interest in the subject matter or materials discussed in this manuscript. 
	\item Ethical Compliance: All procedures performed in studies involving human participants were in accordance with the ethical standards of the institutional and/or national research committee and with the 1964 Helsinki Declaration and its later amendments or comparable ethical standards.
	\item Data availability: The research data supporting this publication are available from the UCI Machine Learning Repository at- \href{https://archive.ics.uci.edu/dataset/45/heart+disease}{https://archive.ics.uci.edu/dataset/45/heart+disease} and the synthetic dataset is available at- \href{https://www.kaggle.com/datasets/mdmahadihassannayem/cleveland-heart-disease-synthetic-dataset/data}{https://www.kaggle.com/datasets/mdmahadihassannayem/cleveland-heart-disease-synthetic-dataset/data}. 
	\item Author contribution: Nayem.M and Biswas.S contributed to the design and implementation of the research. Nayem.M developed the proposed methods. Nayem.M wrote the codes in R and analyzed the results. Nayem.M wrote the manuscript. Additionally, Nayem.M managed the data and interpreted the findings. Biswas.S Conceived the original idea and supervised the project. Biswas.S also critically revised the manuscript. Both authors discussed the results, contributed to the final manuscript, and approved the final version.
	
\end{itemize}

\bibliography{References}

\end{document}